\documentclass[]{spie}  %>>> use for US letter paper
\usepackage{amsmath,amsfonts,amssymb}
\usepackage{graphicx}
\usepackage[colorlinks=true, allcolors=blue]{hyperref}
\usepackage{color}

\title{Review of high-contrast imaging systems for current and future ground-based and space-based telescopes II.\\ Common path wavefront sensing/control and Coherent Differential Imaging}

\author[a]{Nemanja Jovanovic}
\author[b]{Olivier Absil}
\author[c]{Pierre Baudoz}
\author[d]{Mathilde Beaulieu}
\author[e]{Michael Bottom}
\author[e]{Eric Cady}
\author[b]{Brunella Carlomagno}
\author[f]{Alexis Carlotti}
\author[g]{David Doelman}
\author[h]{Kevin Fogarty}
\author[c]{\\Rapha\"{e}l Galicher}
\author[i,j,k]{Olivier Guyon}
\author[g]{Sebastiaan Haffert}
\author[c]{Elsa Huby}
\author[e]{Jeffrey Jewell}
\author[g]{Christoph Keller}
\author[g]{Matthew A. Kenworthy}
\author[j,l]{Justin Knight}
\author[m]{Jonas K\"uhn}
\author[j,l]{Kelsey Miller}
\author[h,n]{Johan Mazoyer}
\author[d]{Mamadou N'Diaye}
\author[g]{Emiel Por}
\author[h]{Laurent Pueyo}
\author[e]{A J Eldorado Riggs}
\author[a]{Garreth Ruane}
\author[o]{Dan Sirbu}
\author[g]{Frans Snik}
\author[e]{J. Kent Wallace}
\author[g]{Michael Wilby}
\author[p]{Marie Ygouf}
\affil[a]{California Institute of Technology, 1200 E. California Blvd., Pasadena, CA 91125, USA}
\affil[b]{Space sciences, Technologies, and Astrophysics Research (STAR) Institute, University of Li\`ege, 19C all\'ee du Six Ao\^ut, B-4000 Li\`ege, Belgium}
\affil[c]{LESIA, Observatoire de Paris, PSL Research University, CNRS, Sorbonne Universit\'es, Univ. Paris Diderot, UPMC Univ. Paris 06, Sorbonne Paris Cit\'e, 5 place Jules Janssen, 92190 Meudon, France}
\affil[d]{Universit\'e Cote d'Azur, Observatoire de la Cote d'Azur, CNRS, Laboratoire Lagrange, Parc Valrose, F-06108 Nice, France}
\affil[e]{Jet Propulsion Laboratory, California Institute of Technology, 4800 Oak Grove Drive, Pasadena, CA 91109 USA}
\affil[f]{CNRS, Institut de Planétologie et d'Astrophysique de Grenoble (IPAG), F-38000, Grenoble, France}
\affil[g]{Leiden Observatory, Leiden University, P.O. Box 9513, 2300 RA Leiden, The Netherlands}
\affil[h]{Space Telescope Science Institute, 3700 San Martin Drive, 21218 Baltimore MD, USA}
\affil[i]{Astrobiology Center, National Institutes of Natural Sciences, 2-21-1 Osawa, Mitaka, Tokyo, JAPAN}
\affil[j]{Steward Observatory, University of Arizona, Tucson, AZ 85721, USA}
\affil[k]{National Astronomical Observatory of Japan, Subaru Telescope, National Institutes of Natural Sciences, Hilo, HI 96720, USA}
\affil[l]{College of Optical Sciences, University of Arizona, 1630 E University Blvd, Tucson, AZ 85719, USA}
\affil[m]{Institute for Particle Physics and Astrophysics, ETH Zurich, Wolfgang-Pauli-Str. 27, CH-8093 Zurich, Switzerland}
\affil[n]{Johns Hopkins University, Zanvyl Krieger School of Arts and Sciences, Department of Physics and Astronomy, Bloomberg Center for Physics and Astronomy, 3400 North Charles Street, Baltimore, MD 21218, USA}
\affil[o]{NASA Ames Research Center, Moffett Field, Mountain View, CA, 94035, USA}
\affil[p]{IPAC, Caltech, 1200 E. California Blvd., Pasadena, CA 91125, USA}

\authorinfo{Further author information: (Send correspondence to Nemanja Jovanovic)\\Nemanja Jovanovic: E-mail:  jovanovic.nem@gmail.com, Telephone: +1 626 395 1214}

\begin{document} 
\maketitle

\begin{abstract}
The Optimal Optical Coronagraph (OOC) Workshop held at the Lorentz Center in September 2017 in Leiden, the Netherlands, gathered a diverse group of 25 researchers working on exoplanet instrumentation to stimulate the emergence and sharing of new ideas. In this second installment of a series of three papers summarizing the outcomes of the OOC workshop (see also~\citenum{ruane2018,snik2018}), we present an overview of common path wavefront sensing/control and Coherent Differential Imaging techniques, highlight the latest results, and expose their relative strengths and weaknesses. We layout critical milestones for the field with the aim of enhancing future ground/space based high contrast imaging platforms. Techniques like these will help to bridge the daunting contrast gap required to image a terrestrial planet in the zone where it can retain liquid water, in reflected light around a G type star from space. 
\end{abstract}

% Include a list of keywords after the abstract 

%Things to add to the final paper
%common path wavefront sensing through fibers - Jorge, Nikita, Dimitri, Yin Zi, SCAR
%LOWFS with a lantern
%add LOWFS
%table of instruments
%link to the new website
%Add better lessons learned, tip/tilt tracking while doing CPWFSing
%A comparison of techniques. Which ones works in what case and why. 

%%%%%%%%%%%%%%%%%%%%%%%%%%%%%%%%%%%%%%%%%%%%%%%%%%%%%%%%%%%%%
%%%-----------------------Intro---------------------------%%%
%%%%%%%%%%%%%%%%%%%%%%%%%%%%%%%%%%%%%%%%%%%%%%%%%%%%%%%%%%%%%
\section{INTRODUCTION}
\label{sec:intro}  % \label{} allows reference to this section
While conventional wavefront sensors (WFSs) measure aberrations using light extracted ahead of the coronagraph optics (usually by a beam splitter), common-path WFSs rely on either exploiting the post-coronagraph focal plane image or the light diffracted by the focal plane mask to infer wavefront aberrations up to the coronagraphic focal plane. In this case the science camera is most commonly used for wavefront sensing. The approach offers two significant advantages: (1) it is free of optical non-common path errors and the measurement is performed at the exact same wavelength as science acquisition and (2) it is highly sensitive, efficiently converting wavefront errors into intensity modulation. 

One challenge is that it is not straightforward to recover wavefront measurements from focal plane intensity alone. Many approaches have been demonstrated to break this degeneracy - all of them relying on coherent mixing with the bright on-axis starlight (see Table~\ref{tab:cpwfs}). The techniques fall into two major categories, namely those that utilize spatial mixing and those that rely on temporal mixing. Spatial mixing refers to techniques that rely on either the continuous interference between the speckle field and a deliberate and permanent probe (e.g. Modal wavefront sensor, self-coherent camera) or the correlation between various points in the complex field about the point-spread function (PSF) (e.g. linear dark field control, Kernel phase). On the other hand, temporal modulation requires the injection of known perturbations, typically with a deformable mirror (DM) or a tip/tilt mirror and the subsequent tracking of the response of the system as a function of time (speckle nulling, pairwise probing, COFFEE, MEDUSAE, phase retrieval). Owing to the fact that some of the science images collected are during probing, the duty cycle of the class of temporal mixing techniques is $<$100\%.  

\begin{table}[ht]
\caption{A list of the various common-path WFSing techniques. LDFC - Linear dark field control, MEDUSAE - Multispectral Exoplanet Detection Using Simultaneous Aberration Estimation, QACITS - Quadrant Analysis of Coronagraphic Images for Tip-tilt Sensing, SCC - Self coherent camera, COFFEE - COronagraphic Focal-plane wave-Front Estimation for Exoplanet detection.} 
\label{tab:cpwfs}
\begin{center}       
\begin{tabular}{|l|c|c|c|c|} %% this creates two columns
%% |l|l| to left justify each column entry
%% |c|c| to center each column entry
%% use of \rule[]{}{} below opens up each row
\hline
\rule[-1ex]{0pt}{3.5ex}  \textbf{Technique} & \textbf{Modulation used} & \textbf{Real time} & \textbf{Science duty cycle} &\textbf{Section}\\
\hline
\hline
\rule[-1ex]{0pt}{3.5ex}  Modal WFS~\citenum{wilby2017}     & no        & yes       & 100\% & \ref{subsec:cMWS} \\
\hline
\rule[-1ex]{0pt}{3.5ex}  LDFC~\citenum{guyon2016,kelsey2017}       & no        & yes       & 100\%  &\ref{subsec:LDFC}\\
\hline
\rule[-1ex]{0pt}{3.5ex}  MEDUSAE~\citenum{marie2013}               & no        & no        & $<100$\%  &\ref{subsec:coffee}\\
\hline
\rule[-1ex]{0pt}{3.5ex}  COFFEE~\citenum{paul2014}                  & yes       & no        & $<100$\%  &\ref{subsec:coffee}\\
\hline
\rule[-1ex]{0pt}{3.5ex}  QACITS~\citenum{mas2012,huby2015}         & no        & yes       & 100\% &\ref{subsec:qacits}\\
\hline
\rule[-1ex]{0pt}{3.5ex}  SCC~\citenum{baudoz2006}            & yes       & yes       & 100\%  &\ref{subsec:scc}\\
\hline
\rule[-1ex]{0pt}{3.5ex}  Pairwise probing~\citenum{giveon2011} & yes     & no        & $<100$\%  &\ref{subsec:pairwise}\\
\hline
\rule[-1ex]{0pt}{3.5ex}  Speckle nulling~\citenum{borde2006, martinache2014} & yes      & yes       & $<100$\%  &\ref{subsec:spec_null}\\
\hline
\rule[-1ex]{0pt}{3.5ex}  Phase retrieval~\citenum{fienup1982}       & yes      & no        & $<100$\%  &\ref{subsec:ph_retriev}\\
\hline
\rule[-1ex]{0pt}{3.5ex}  Kernel phase~\citenum{martinache2010,martinache2013}      & yes       & yes       & $<100$\%  &\ref{subsec:kernel}\\
\hline
\rule[-1ex]{0pt}{3.5ex}  Phase shifting interferometry~\citenum{bottom2017}      & no       & yes       & $100$\%  &\ref{subsec:psi1}\\
\hline
\rule[-1ex]{0pt}{3.5ex}  Phase sorting interferometry~\citenum{Codona13, frazin2013}      & no       & yes       & $100$\%  &\ref{subsec:psi2}\\
\hline
\end{tabular}
\end{center}
\end{table} 

%\rule[-1ex]{0pt}{3.5ex}  Zernike WFS~\citenum{Wallace2011,Ndiaye2016}      & yes       & no       & $<100$\%  &\\

As originally proposed and implemented, the goal of common-path WFSs was solely to measure coherent light in the focal plane, so that a control loop could drive this residual light (coronagraph leakage) toward zero (usually by DM actuation). The same signals that lead to wavefront measurements can also be processed to decompose the focal plane image into both a coherent component and an incoherent component. The coherent measurement drives the high contrast control loop, while the incoherent component contains the astrophysical signal, and should therefore be collected. The approach of modulating the speckle field to determine the incoherent/coherent parts of the field is referred to as coherent differential imaging (CDI) and is typically undertaken in post-processing (similar to angular differential imaging or spectral differential imaging). 

Owing to the finite temporal bandwidth of the wavefront control loop versus the coherence time of the atmosphere, the limited sensitivity of a WFS, detector noise, chromatic errors and so on, real time correction using a common-path WFS can only improve the contrast over a conventional WFS by so much. Post-processing via CDI methods will be critical to enhance the contrast beyond these limitations and bridge the extreme contrast of $10^{10}$ between a G type star like our sun and a terrestrial planet orbiting it in the habitable zone.

In this paper we review some of the common-path WFSing/CDI techniques and outline what has been achieved to date. We list some lessons learned and set out milestones for the field to be used as guidelines for development to enable imaging of terrestrial exoplanets in the future. 

%%%%%%%%%%%%%%%%%%%%%%%%%%%%%%%%%%%%%%%%%%%%%%%%%%%%%%%%%%%%%%%%%%%%%%%%%%
%%%-----------------------Technique overview---------------------------%%%
%%%%%%%%%%%%%%%%%%%%%%%%%%%%%%%%%%%%%%%%%%%%%%%%%%%%%%%%%%%%%%%%%%%%%%%%%%
\section{Overview of common-path wavefront sensing techniques and state of the field}
This section steps through a selection of promising common-path WFSing techniques, outlines the basic concept behind each and provides an overview of their maturity and success thus far. 

%%%%%%%%%%%%%%%%%%%%%%%%%%%%%%%%%%%%%%%%%%%%%%%%%%%%%%%%%%%%%%%%%%%%%%%%%%
\subsection{Coronagraphic Modal Wavefront Sensor (cMWS)}
\label{subsec:cMWS}
In addition to creating dark holes in the stellar PSF, pupil-plane coronagraphs can also be used to encode wavefront-sensing information directly in the science focal plane. In the case of an Apodizing Phase Plate (APP) coronagraph~\citenum{Codona06,Kenworthy07}, phase-only holograms can be multiplexed with the APP pattern to produce pairs of modified copies of the central PSF at arbitrary locations in the focal plane, of arbitrary relative power, and containing independent wavefront biases with respect to the central science image.

\begin{figure} [t!]
   \begin{center}
   \begin{tabular}{c} %% tabular useful for creating an array of images 
   \includegraphics[width=0.95\textwidth]{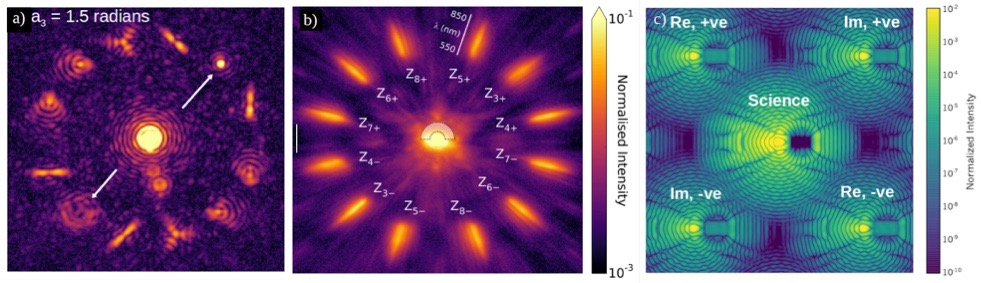}
   \end{tabular}
   \end{center}
   \caption[example] 
%>>>> use \label inside caption to get Fig. number with \ref{}
   { \label{fig:mWFS} 
(Left) Laboratory image of a 6-mode Zernike cMWS with 1.5 radians RMS of focus error, showing asymmetrical response of the corresponding focus PSF copies. (Middle) On-sky demonstration of a cMWS at the William Herschel telescope using the same sensing modes, for 20\% bandwidth. (Right) Numerical simulation of a PSF for holographic electric field conjugation, with four copies each containing a different electric field probe in the APP dark hole (Figures adapted from~\citenum{por2016,wilby2017}).}
\end{figure}

The coronagraphic Modal Wavefront Sensor (cMWS) uses this holographic technique to create two oppositely biased PSF copies for each wavefront mode to be sensed~\citenum{wilby2017} (Fig.~\ref{fig:mWFS}, left panel).
The normalized difference between the core intensities of corresponding PSF copies is then directly related to the mode amplitude, allowing fast retrieval of the wavefront coefficients at the focal plane using two intensity measurements per mode. 

The creation of each holographic PSF copy requires a small reduction to science throughput, which for the cMWS is typically chosen to be 1\% per mode. This, coupled with limitations on focal-plane detector space, typically makes it impractical to sense more than 20-30 modes. However, a free choice of (orthogonal) sensing mode basis means that the cMWS is not necessarily limited to only low-order aberrations: it is equally possible to use a tailored basis encompassing all spatial frequencies which correspond to a small dark hole a few square $\lambda$/D in size.

A cMWS prototype was first implemented on-sky at the William Herschel telescope (WHT) in 2015~\citenum{wilby2017}, with improved designs included as an open-loop subsystem of the LEXI instrument with first light in June 2016~\citenum{haffert2016,wilby2016}. These first tests demonstrated that the cMWS is capable of the real-time sensing of low-order Zernike aberrations injected via reference offsets on the closed-loop AO system, and is able to operate over arbitrarily wide bandwidths (Fig.~\ref{fig:mWFS}, middle panel). Most recently, successful on-sky closed-loop operation was achieved during the second LEXI observing run at the WHT in November 2017~\citenum{haffert2018}, using optimized cMWS designs manufactured as liquid crystal phase plates.

This holographic spatial multiplexing approach can also be extended to operate a variety of temporally-modulated focal-plane WFS techniques in a spatially-modulated manner, by using one or more secondary PSF copies instead of the science PSF. This allows for a 100\% duty cycle at the cost of some science throughput, making real-time operation feasible. For example, phase diversity or curvature sensing can be performed by adding a parabolic phase term to a single PSF copy, while electric field conjugation can be achieved by adding well-defined electric field probes to the dark hole of four or more PSF copies~\citenum{por2016} (Fig.~\ref{fig:mWFS}, right panel).

%%%%%%%%%%%%%%%%%%%%%%%%%%%%%%%%%%%%%%%%%%%%%%%%%%%%%%%%%%%%%%%%%%%%%%%%%%
\subsection{Linear Dark Field Control (LDFC)}
\label{subsec:LDFC}
Linear dark field control (LDFC) provides a relative measurement of the change in the field based on the change in the intensity of bright speckles outside the dark hole so that the contrast in the dark hole can be maintained. It is not an absolute wavefront measurement technique and so can not be used to dig the dark hole in the first instance.   

LDFC relies on two key properties, namely the linear response of the intensity of a bright speckle as a function of the amplitude of the input wavefront perturbation and that there are a subset of aberrations that will effect both the bright field and light in the dark hole simultaneously. Therefore, by examining only the bright speckle field, the contrast in a dark hole next to the PSF core can be controlled/maintained. In this way, aberrations that effect the dark hole can be sensed and canceled, which maintains the dark hole at its initial deep contrast state. 

\begin{figure} [ht]
   \begin{center}
   \begin{tabular}{c} %% tabular useful for creating an array of images 
   \includegraphics[width=0.70\textwidth]{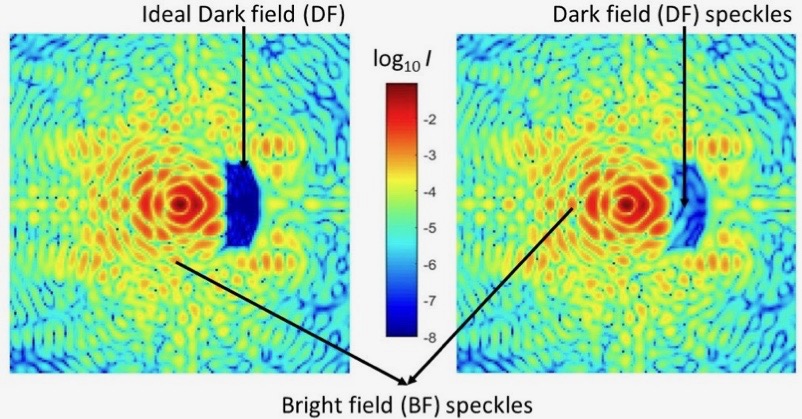}
   \end{tabular}
   \end{center}
   \caption[example] 
%>>>> use \label inside caption to get Fig. number with \ref{}
   { \label{fig:LDFC} 
(Left) Spatial LDFC: 3.5 $\lambda/D$ x 8.5 $\lambda/D$ dark hole/dark field (DF) created in numerical simulation using conventional electric field conjugation (EFC). (Right) Wavefront aberrations produce speckles in the bright field (BF) and the DF which degrade the DF. The change in intensity between the aberrated BF (Right) and the ideal BF (Left) is used to measure and cancel the speckles in the DF.}
\end{figure}

LDFC comes in two flavors: spatial~\citenum{kelsey2017} and spectral~\citenum{guyon2016}. Spatial LDFC, shown in Fig.~\ref{fig:LDFC}, uses information from speckles located spatially outside of the dark hole at the science wavelength(s) whereas spectral LDFC senses speckles that occur spatially within the dark hole but spectrally outside of the control bandwidth. 
In principle, the two can be run simultaneously to compensate for each technique’s null space. LDFC can be deployed as a deep contrast stabilization technique for any single-sided dark hole; it cannot be implemented on $360^{\circ}$ dark holes since it relies on access to the same spatial frequencies in the bright field opposite the dark hole. Since it does not rely on modulation or light from the very faint dark hole, LDFC can return the dark hole to its initial deep contrast much faster than modulation-based techniques that probe the field within the dark hole.      

Thus far both spatial and spectral LDFC have been simulated and shown to work in principle. While spectral LDFC has not been tested experimentally to date, spatial LDFC was recently demonstrated on the SCExAO instrument (Guyon et al. in prep.). In this demonstration a dark hole was created through speckle nulling (discussed in section~\ref{subsec:spec_null}, and then artificial speckles were injected into the hole and shown to be suppressed by the LDFC loop. There are also plans to validate the technique on the University of Arizona’s Wavefront Control Testbed with the two single-sided dark holes created by a vector APP coronagraph. Experiments are planned to implement spectral LDFC in a slow, closed-loop fashion using data from the CHARIS IFS~\citenum{groff2017} behind the SCExAO instrument~\citenum{jovanovic2015}.

%%%%%%%%%%%%%%%%%%%%%%%%%%%%%%%%%%%%%%%%%%%%%%%%%%%%%%%%%%%%%%%%%%%%%%%%%%
\subsection{MEDUSAE/COFFEE}
\label{subsec:coffee}
\begin{figure} [hb]
   \begin{center}
   \begin{tabular}{c} %% tabular useful for creating an array of images 
   \includegraphics[width=0.95\textwidth]{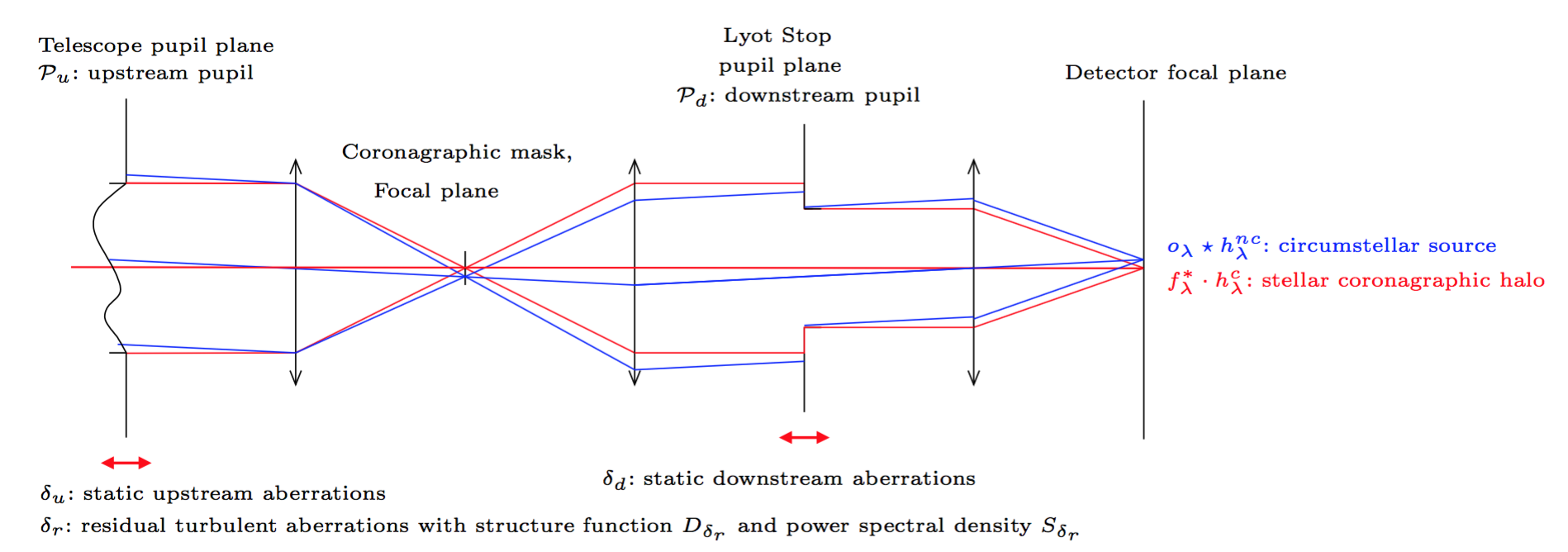}
   \end{tabular}
   \end{center}
   \caption[example] 
%>>>> use \label inside caption to get Fig. number with \ref{}
   { \label{fig:COFFEE} 
Optical scheme of the coronagraphic imager model used in the MEDUSAE and COFFEE algorithms (Fig.~2 of~\citenum{marie2013} adapted from~\citenum{Sauvage2010}). In its most current implementation, MEDUSAE estimates jointly the quasi-static instrumental aberrations located upstream of the coronagraph $\delta_{u}$ and the circumstellar source $O_{\lambda}$ directly from multispectral science images. In the most current implementation, COFFEE jointly estimates the upstream electric field and the quasi-static instrumental aberrations downstream $\delta_{d}$ of the coronagraph using DM probes and a realistic model of a coronagraph.}
\end{figure}

The MEDUSAE and COFFEE algorithms are two flavors of a family of techniques based on the use of a model of a coronagraphic imaging system that links the speckle field in the focal plane to some parameters of interest (see Fig.~\ref{fig:COFFEE} for details). These parameters are then estimated using an inverse problem approach in a Bayesian framework. The two techniques have been developed to be complementary and eventually used in synergy (COFFEE for calibration and MEDUSAE for post-processing).  

\begin{sloppypar}
The MEDUSAE (Multispectral Exoplanet Detection Using Simultaneous Aberration Estimation) post-processing technique~\citenum{marie2013} jointly estimates the instrumental aberrations and the object of interest, i.e. the circumstellar environment, in order to separate these two contributions. To do so, it makes use of the Simultaneous Differential Imaging (SDI) strategy by exploiting the wavelength diversity offered by an integral field unit or a dual band imager. The inversion algorithm is based on a maximum a posteriori estimator, which measures a regularized discrepancy between the multispectral data and the imaging model. It is then possible to estimate the aberration map and the circumstellar environment at each wavelength — and thus their spectra — directly from the science image. Despite the fact that MEDUSAE has been tested and validated on simulated SPHERE-IFS images for the purpose of post-processing~\citenum{marie2013}, it has the potential to be used in real time in the context of wavefront sensing and control as well. Near-term prospects are to: 1) validate the technique on SPHERE data for post-processing~\citenum{Cantalloube17} and 2) implement a non-coronagraphic model in order to validate it on the NIRSpec IFU on JWST for post-processing~\citenum{Ygouf17}.
\end{sloppypar}

COFFEE (COronagraphic Focal-plane wave-Front Estimation for Exoplanet detection)~\citenum{paul2014} is a coronagraphic phase diversity technique that estimates the instrumental aberrations within a calibration framework. It uses the same model and core routines as MEDUSAE with a few differences in the way it is implemented. For example, it utilizes DM probes for diversity rather than the wavelength diversity used by MEDUSAE. Also, it makes use of a Matrix Fourier Transform (MTF)~\citenum{Soummer2007} to simulate a more realistic coronagraph, which is necessary for its implementation on a testbed. The COFFEE principle was demonstrated in the laboratory in Marseille in France on the MITHIC testbed. It was used to estimate the phase aberrations upstream and downstream of a focal plane mask coronagraph from the science image. The minimization of these phase aberrations thanks to a DM led to a $3\times10^{-8}$ dark hole at $5\lambda/D$ separation~\citenum{Paul2013}. Future prospects are to test and validate the technique on several other testbeds including the THD testbed.

%%%%%%%%%%%%%%%%%%%%%%%%%%%%%%%%%%%%%%%%%%%%%%%%%%%%%%%%%%%%%%%%%%%%%%%%%%
\subsection{Quadrant Analysis of Coronagraphic Images for Tip-tilt Sensing (QACITS)} 
\label{subsec:qacits}
\begin{figure} [hb]
   \begin{center}
   \begin{tabular}{c} %% tabular useful for creating an array of images 
   \includegraphics[width=0.60\textwidth]{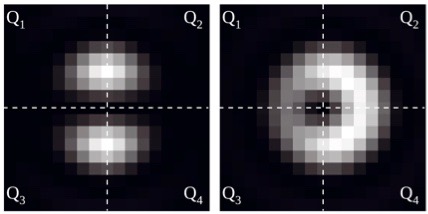}
   \end{tabular}
   \end{center}
   \caption[example] 
%>>>> use \label inside caption to get Fig. number with \ref{}
   { \label{fig:QACITS}
Quadrant analysis of the coronagraphic image in the case of a non obstructed pupil (left) for the four quadrant phase phase mask and (right) for the vortex phase mask. The tip-tilt amplitude is $0.2~\lambda/D$ and the image width is $4~\lambda/D$.}
\end{figure}

The QACITS is a tip-tilt sensing technique based on the analysis of the flux asymmetry of the coronagraphic image that appears when the star image is not perfectly centered on the coronagraphic focal plane mask, as shown in Fig.\ref{fig:QACITS}. This technique is designed to work only with a focal-plane phase mask upstream. The measured differential flux intensity (like in a quad-cell position sensor) within the central area ($< 3\lambda/D$) is indeed related to the amplitude of the pointing error affecting the beam incident on the focal plane phase mask. Models have been derived for the four quadrant phase mask~\citenum{mas2012} and for the vortex phase mask~\citenum{huby2015}. In case of an obstructed pupil, the image analysis has to be done in separate areas in order to disentangle the contribution of light diffracted by the central obstruction~\citenum{huby2017}: the flux is measured in more than four areas.

The QACITS algorithm is routinely used on-sky ~\citenum{huby2017} with the vortex coronagraph on the Keck/NIRC2 infrared camera (L and M bands). The control loop is typically run at a frequency of 0.02~Hz to correct for slow drifts and stabilize the position of the star image behind the coronagraphic focal plane mask down to 2.4~mas RMS.

%%%%%%%%%%%%%%%%%%%%%%%%%%%%%%%%%%%%%%%%%%%%%%%%%%%%%%%%%%%%%%%%%%%%%%%%%%
\subsection{Self Coherent Camera (SCC)}
\label{subsec:scc}
\begin{figure} [ht]
   \begin{center}
   \begin{tabular}{c} %% tabular useful for creating an array of images 
   \includegraphics[width=0.70\textwidth]{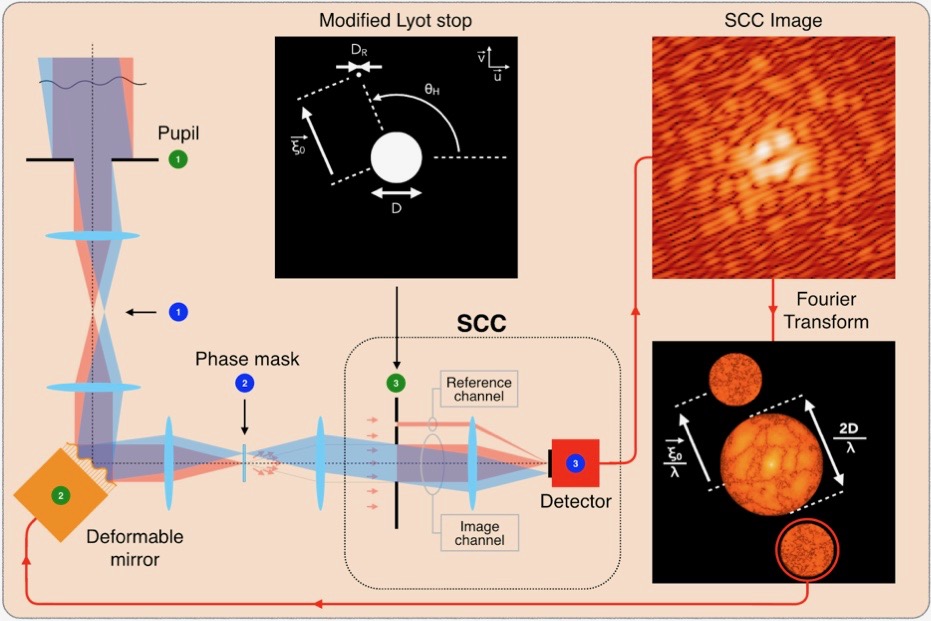}
   \end{tabular}
   \end{center}
   \caption[example] 
%>>>> use \label inside caption to get Fig. number with \ref{}
   { \label{fig:SCC} 
The principle of the SCC. The stellar light (red beam) hits a DM and is focused on a coronagraphic focal plane mask. Most of the light is diffracted outside the geometrical pupil downstream and is blocked by a Lyot stop. The coronagraphic leak is transmitted at the Lyot stop and creates a speckle image on the detector. The small hole carefully positioned in the Lyot stop selects the reference beam which is interfered with the light leaked by the coronagraph which spatially encodes speckles (top right panel). It is then possible to retrieve the speckle electric field using a Fourier transform (bottom right panel).}
\end{figure}

The self-coherent camera (SCC) concept utilizes a small off-axis “reference” hole in the otherwise opaque part of a Lyot stop of a coronagraph, to transmit some of the light rejected by the coronagraph. The downstream optics must be sufficiently large to capture both the primary coronagraphic leakage inside the geometric pupil and the reference beam. By forming an overlapping image with the two beams, Fizeau interference fringes are created across the speckle field in the science image as seen in Fig.~\ref{fig:SCC}. 
In this way, the electric field in the science image is spatially modulated and can be directly retrieved. Only one image of the science channel is therefore necessary to encode the speckles and estimate the electric field that is to be minimized to create a dark hole: the correction is done at the rate of the science image recording. One can add other “reference” holes in the Lyot pupil to make the SCC work in polychromatic light~\citenum{delorme2016b}. 

\begin{figure} [ht]
   \begin{center}
   \begin{tabular}{c} %% tabular useful for creating an array of images 
   \includegraphics[width=0.70\textwidth]{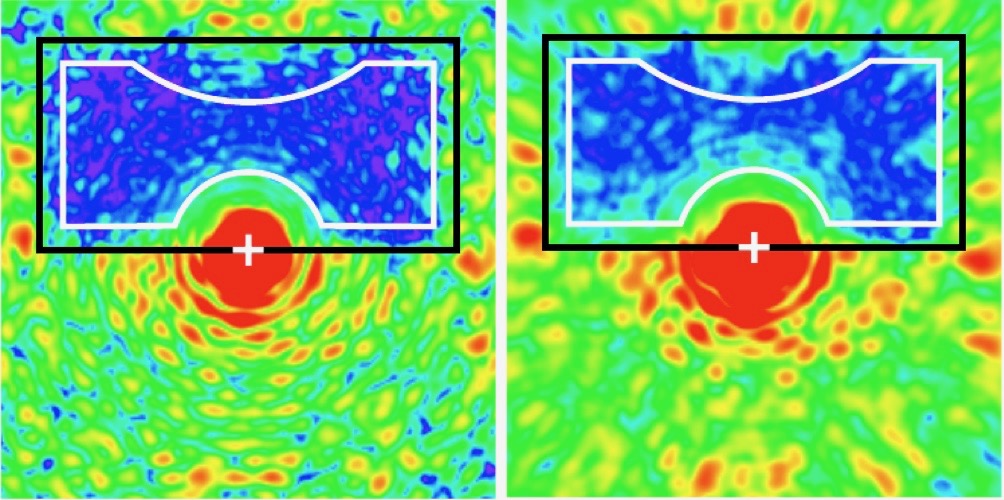}
   \end{tabular}
   \end{center}
   \caption[example] 
%>>>> use \label inside caption to get Fig. number with \ref{}
   { \label{fig:SCC_DH} 
SCC Laboratory results. (Left) Half dark hole (13x27 $\lambda/D$) obtained in monochromatic (650nm) and (Right) broadband (12,5\% bandwidth in visible) light using the SCC as a focal plane wavefront sensor. The bottom of the dark hole (blue and purple colors) is $10^{-8}$ contrast~\citenum{delorme2016}.}
\end{figure}

The SCC images can be used for both CDI~\citenum{baudoz2006, galicher2007,galicher2008} as well as for focal plane wavefront sensing~\citenum{galicher2008, baudoz2010, baudoz2010b,  galicher2010, mas2010, baudoz2012, mas2012b, mazoyer2012}. Wavefront sensing with the SCC has been demonstrated in both numerical simulations and in the laboratory. It was used to reach a contrast at the $10^{-8}$ level routinely on the THD bench (space-like conditions but no vacuum). It was tested with numerous coronagraph architectures including the four quadrant phase mask~\citenum{mazoyer2013, mazoyer2013b}, the multi four quadrant phase mask~\citenum{galicher2014}, the apodized dual zone phase mask~\citenum{delorme2016, delorme2016b}, the eight-octant phase mask, the vector vortex, and the six level phase mask~\citenum{baudoz2018}. Results in monochromatic light~\citenum{mazoyer2014} and 12.5\% bandwidth~\citenum{delorme2016} have been obtained on Meudon THD testbed and both reached a contrast level of $10^{-8}$ before any CDI calibration (see Fig.~\ref{fig:SCC_DH}).

\begin{figure} [ht]
   \begin{center}
   \begin{tabular}{c} %% tabular useful for creating an array of images 
   \includegraphics[width=0.35\textwidth]{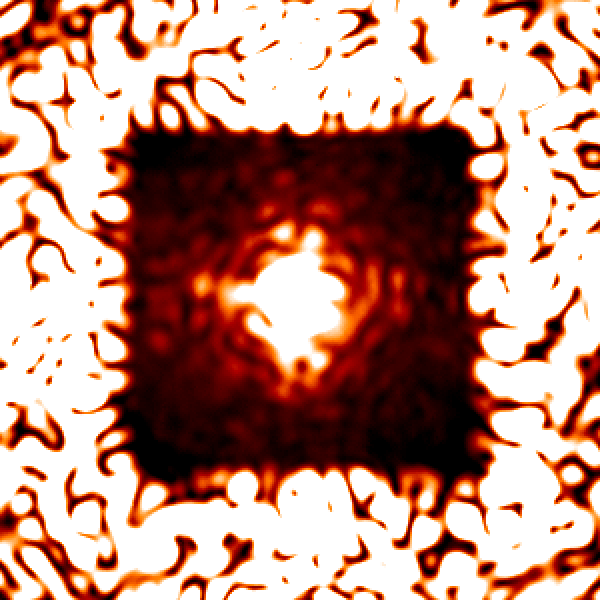}
     \includegraphics[width=0.35\textwidth]{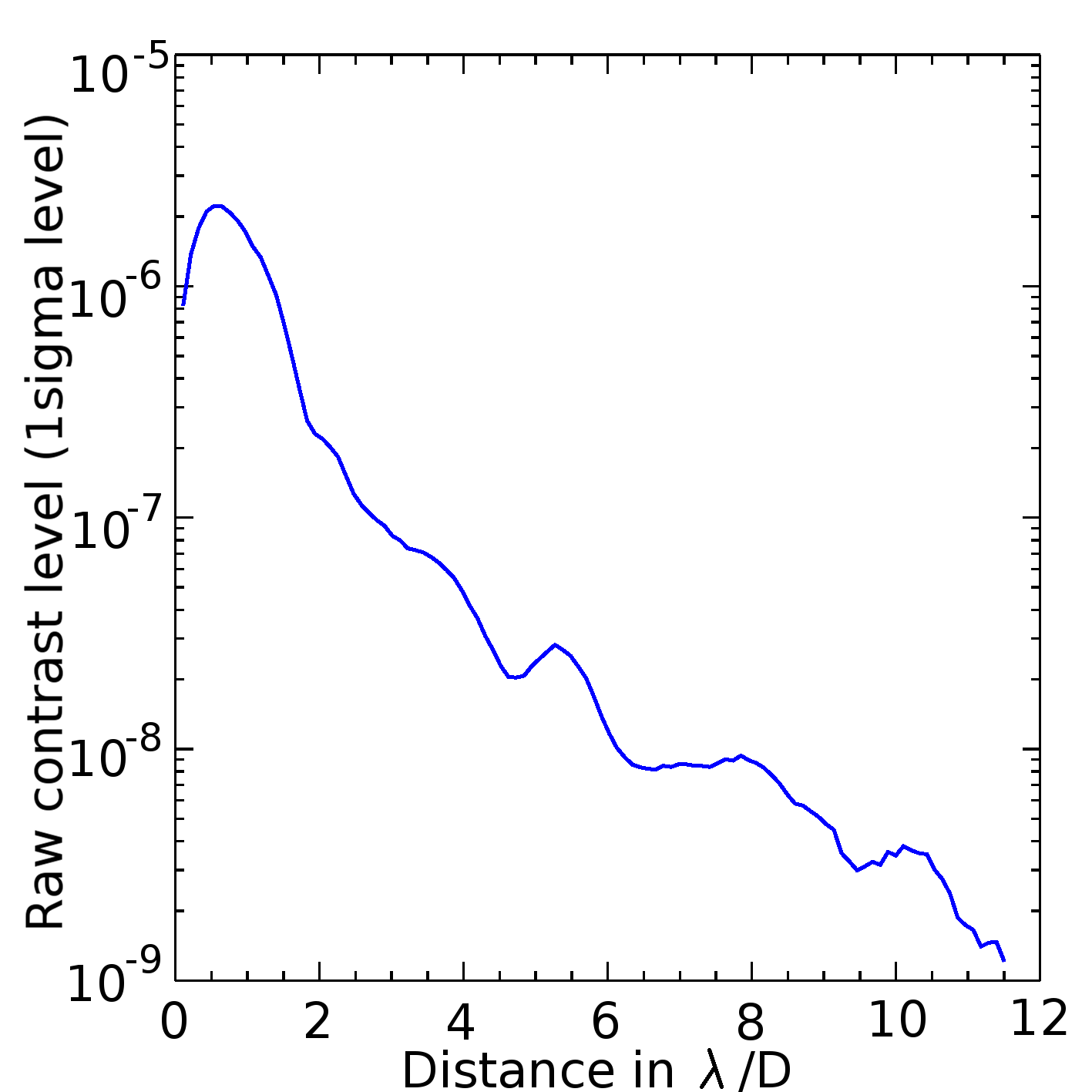}
   \end{tabular}
   \end{center}
   \caption[example] 
%>>>> use \label inside caption to get Fig. number with \ref{}
   { \label{fig:SCC_DH_2DM} 
SCC Laboratory results. (Left) Full dark hole (17x17 $\lambda/D$) obtained in monochromatic light (785nm) using the SCC as a focal plane wavefront sensor and two deformable mirrors. No CDI is applied. (Right) Normalized azimuthal standard deviation associated to the image~\citenum{baudoz2018b}.}
\end{figure}

Recently, the SCC was also used to control two DMs in cascade in order to correct both phase and amplitude aberrations, resulting in a $360^{\circ}$ dark hole with a $<10^{-8}$ contrast level (figure~\ref{fig:SCC_DH_2DM}) \citenum{baudoz2018b}. Finally, the SCC is currently installed at the Palomar Observatory on the stellar double coronagraph instrument for on-sky testing. Gerard et al. (2018) recently proposed a modified focal plane mask for the SCC where fringes would be detected in millisecond-timescale exposures, allowing focal plane wavefront sensing and control of both atmospheric and quasi-static speckles~\citenum{gerard2018}.

%%%%%%%%%%%%%%%%%%%%%%%%%%%%%%%%%%%%%%%%%%%%%%%%%%%%%%%%%%%%%%%%%%%%%%%%%%
\subsection{Pairwise probing}
\label{subsec:pairwise}
Pairwise probing uses pairs of DM motions (“probes”), of the same magnitude but opposite sign, to modulate the electric field in defined regions of the science focal plane. Pairs of probes are used so that sums and differences between images cancel out common-mode terms, such as probe intensity or probe/star cross terms. By combining two or more pairs of probes with an unprobed image and a model of the coronagraphic system, the complex electric field of the starlight and the “incoherent” residual may be simultaneously estimated across large focal-plane areas. In pairwise probing, the incoherent residual includes both terms which are truly incoherent with the starlight, such as planet and exozodiacal light, and estimation error in the coherent signal, which complicates subsequent analysis; the estimation error is generally the dominant term until the coherent level is $3$--$4\times$ the incoherent floor.

A common choice for probes is to place a pair of sinc functions on the DM, one oriented along the x-axis and one along the y-axis; these are then multiplied by a sine wave whose phase may be modulated. The sinc phase modulation creates a rectangular region, while the sine makes two copies of this region on either side of the PSF peak. Modulating the sine wave shifts the relative phase of the two rectangles, beating the probe field against the remaining speckle field. Further extensions to this method include the use of several probes, which increase the robustness to noise at the expense of overhead, the use of an estimator such as a Kalman filter to replace one of the probes while incorporating previous knowledge about the testbed state, and the use of holographic gratings to introduce probe images into the focal plane concurrently with the science exposure.

For the primary description of the technique, including the mathematical description of the underlying algorithm, see~\citenum{giveon2011}. Discussion of the role of estimation error in the incoherent signal can be found in~\citenum{cady2014}, extensions to pairwise probing are covered in~\citenum{riggs2016}, and the introduction of holographically generated probes is discussed in~\citenum{por2016}.

Pairwise probing with sequential DM probes has been demonstrated at several laboratory testbeds, including NASA/JPL, Princeton, and NASA/Ames, with an eye towards space applications. These estimation methods have been combined with correction algorithms such as electric field conjugation (EFC) or stroke minimization to create very-high-contrast dark holes in stabilized environments; estimates of coherent and incoherent components are a byproduct of this correction procedure.   A summary of high-contrast results as of 2013 is given in Table 1 of~\citenum{lawson2013}; in that table, HCIT (NASA/JPL) and ACE (NASA/Ames) use pairwise probing in their estimation.  Most recent results from testing for the WFIRST CGI instrument~\citenum{noecker2016} have yielded $1.6\times10^{-9}$ over $3-9\lambda/D$ in a $10\%$ band with an hybrid Lyot coronagraph~\citenum{seo2017} and $1.1\times10^{-8}$ in an $18\%$ band with a shaped pupil coronagraph~\citenum{cady2017, groff2017_2}.

\begin{figure} [ht]
   \begin{center}
   \begin{tabular}{c} %% tabular useful for creating an array of images 
   \includegraphics[width=0.47\textwidth]{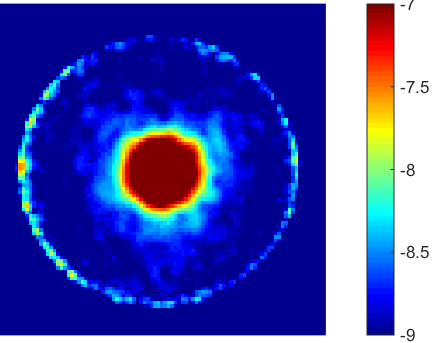}
     \includegraphics[width=0.30\textwidth]{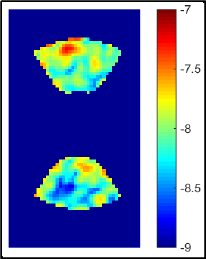}
   \end{tabular}
   \end{center}
   \caption[example] 
%>>>> use \label inside caption to get Fig. number with \ref{}
   { \label{fig:CGI} 
Laboratory results from WFIRST CGI testing at JPL, using pairwise probing for wavefront estimation.  Both are shown on the same base-10-log contrast scale.  \textit{Left.} Best narrow-field-of-view result with an HLC coronagraph ($1.6\times10^{-9}$ contrast $3-9\lambda/D$, $10\%$) \textit{Right.} Best spectroscopy-mode result with an SPC coronagraph ($1.1\times10^{-8}$ contrast $3-9\lambda/D$, $18\%$)
}
\end{figure}

%%%%%%%%%%%%%%%%%%%%%%%%%%%%%%%%%%%%%%%%%%%%%%%%%%%%%%%%%%%%%%%%%%%%%%%%%%
\subsection{Speckle nulling}
\label{subsec:spec_null}
Sinusoidal phase patterns in the pupil plane (which are easily implemented with a DM) correspond to symmetric speckles in the image plane. A small amplitude cosine phase aberration corresponds to symmetric, image plane speckles which are purely imaginary. Likewise, sinusoidal phase aberrations in the pupil plane produce purely real, and antisymmetric speckles in the image plane. Therefore, any speckle in the image plane can be nulled by introducing a sum of sine and cosine phase aberrations in the pupil plane such that the real component of the speckle is nulled by the amplitude of the sine wave, and the imaginary component is nulled by the amplitude of the cosine wave. In practice, since the real and imaginary components of the image plane speckles are unknown, a magnitude and phase of a cosine wave are initially estimated, and applied to the DM. For a given amplitude, the phase of the sinusoidal wave is varied by three steps of $\pi/2$ phase increments, and the speckles brightness is monitored during these phase steps. From these four intensity measurements, it is possible to estimate the real and imaginary parts or the speckle electric field, and the appropriate anti-speckle is applied to reduce the speckles brightness. The process is repeated until the speckles brightness reaches an acceptable threshold. 

This method is powerful in its simplicity. It relies upon no assumption of the system whatsoever. Typically, the method is used to suppress several speckles simultaneously. Because of its simplicity, it has been used for many high-contrast imaging systems. In order to implement the system, the only requirement is to add a cosine phase ripple of a given amplitude and phase to the DM, and to record the resulting science image. The steps can be implemented on the sky, and will be effective depending upon the magnitude of the speckles, and the sample time of the science camera. Of course a larger number of steps during the phase scan can be used to improve the precision of the sinusoidal fit, but the scan time must be taken into account when considering the timescale of changes in the wavefront. Given that at minimum $4$ images are needed with the probes during the phase scan, the maximum science duty cycle (frames that do not have any DM modulation during capture) is $20\%$ during the nulling process. If one intends to use this loop to chase speckles from turbulence, then this duty cycle is fixed. However, if the goal is to null quasi-static speckles only, then the nulling process can be run for a period and once it has converged, the dark hole could be maintained while science exposures are collected at $100\%$ duty cycle (no probes applied). Updates to the dark hole need to be administered as the quasi-statics change with time but this will be on minute-10s of minute timescales. Of course, if CDI techniques are employed in post-processing, then it may be possible to utilize all frames (with and without probes) for the final data reduction. This has not been shown to date. 
 
Speckle nulling was first demonstrated in the laboratory on the HCIT testbed at JPL~\citenum{borde2006}. More recently it has been tested on the SCExAO testbed in both the laboratory~\citenum{martinache2012} and on-sky~\citenum{martinache2014}. The laboratory tests were even conducted with a PIAA coronagraph to demonstrate that the remapping function did not affect the ability to null the speckles~\citenum{martinache2012}. The first on-sky demonstration was limited by the wavefront control provided by the Subaru Telescope facility AO system, AO188 and only a modest improvement of a factor of 2 in the variance was observed over the area of the first Airy ring~\citenum{martinache2014}. In more recent work, this method was shown to convincingly suppress quasi-static speckles over the entire control region of the DM (one side of the image) in the presence of atmospheric speckles~\citenum{jovanovic2015}. A similar approach has been tested on Palomar in the H band and Keck in the L band~\citenum{bottom2016}. The latter included the use of the vortex coronagraph. Unlike in the SCExAO example where the quasi-static speckles were a result of amplitude errors, the Keck demonstration was capable of creating a $360^{\circ}$ dark hole around the PSF. The contrast was greatly improved by applying this technique, but was limited in spatial frequency due to the finite number of actuators on the Keck DM and the slow read out rate of the science camera.      

The H band nulling demonstration on SCExAO was limited by the finite speed and non-negligible read out noise ($\sim120~e^{-}$) of the InGaAs camera used, so it could not chase the atmospheric speckles which set an upper limit to the maximum contrast improvement possible with this technique on-sky. New detector technologies like the SAPHIRA and MKIDS will help alleviate these issues. However, this technique is most effective if it can be applied when operating in the diffraction limited regime for which you need to be able to run this in conjunction with a high order AO system. Inroads into offsetting the probe signals from the pyramid WFS correction in the SCExAO system have been made, but significant work still remains in this area.

%%%%%%%%%%%%%%%%%%%%%%%%%%%%%%%%%%%%%%%%%%%%%%%%%%%%%%%%%%%%%%%%%%%%%%%%%%
\subsection{Phase retrieval}
\label{subsec:ph_retriev}
Phase retrieval methods are regularly used in high-contrast imaging testbeds for calibration purposes. For example, phase retrieval can be used to obtain the DM registration required for testbed models to enable the mid-spatial frequency wavefront control loop. Another common usage of phase retrieval on a testbed is to obtain a new DM flat setting that improves the Strehl ratio and can aid in improving the mid-spatial frequency control loop convergence. Thus a phase retrieval process is almost always used as part of the initial testbed model calibration, and it can also be re-run periodically to update the model.

%The impact of model calibration uncertainty errors was recently estimated for the WFIRST testbed~\citenum{marx2017}. - Nem - this was the first sentence but is seems out of context so I removed it.  - Dan - moved this in the second paragraph in the correct location

There exist a wide variety of different phase retrieval implementations (these families of methods are classified in~\citenum{fienup1982}). The most common is based on a Gerchberg-Saxton iterative process in which a set of focal-plane images are collected. A focal plane image is transformed to a pupil plane, known amplitude and phase constraints are applied before transforming back to the image plane and repeating the process. One problem with the classical Gerchberg-Saxton process is that convergence to the global minimum solution is not guaranteed with convergence plateauing around a local minimum. To improve solution convergence, it is useful to introduce a known diversity such as a defocus term. This defocus term can be implemented in a testbed by translation of the camera along the optical axis or changing the f-number of the focusing optic. Movement of the secondary was used to analyze phase errors on Hubble~\citenum{krist1995}. Phase retrieval is commonly applied as the first step at JPL's HCIT testbed to obtain the DM setting and its registration~\citenum{zhou2017} and reduce model calibration uncertainty errors for WFIRST experiments~\citenum{marx2017}. Phase retrieval can also be used to estimate errors upstream and downstream of a focal plane mask as was done for the EXCEDE experiment~\citenum{sirbu2016}. At the NASA Ames testbed, recent experiments have shown that the DM can be used to generate diversity for phase retrieval without requiring accurate parametric knowledge of the diversity~\citenum{pluzhnik2017}.

%%%%%%%%%%%%%%%%%%%%%%%%%%%%%%%%%%%%%%%%%%%%%%%%%%%%%%%%%%%%%%%%%%%%%%%%%%
\subsection{Kernel phase}
\label{subsec:kernel}
\begin{figure} [ht]
   \begin{center}
   \begin{tabular}{c} %% tabular useful for creating an array of images 
   \includegraphics[width=0.40\textwidth]{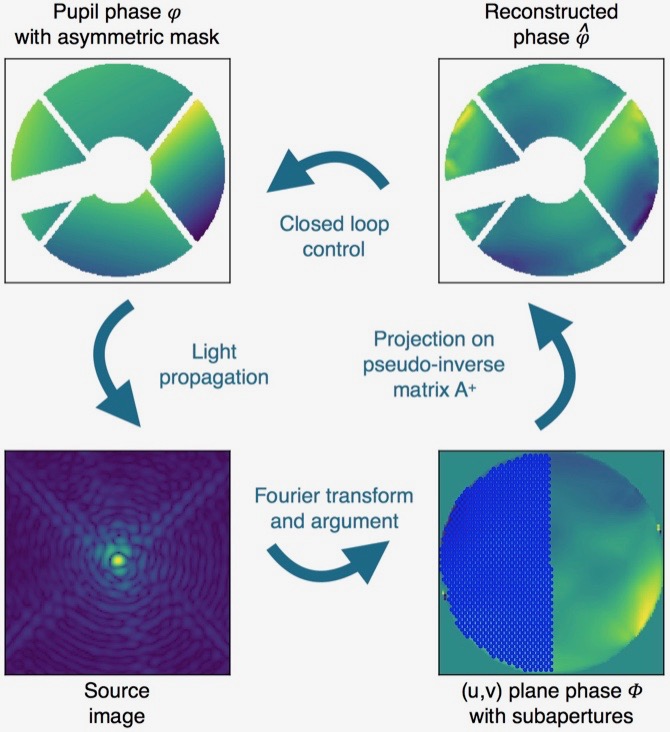}
   \end{tabular}
   \end{center}
   \caption[example] 
%>>>> use \label inside caption to get Fig. number with \ref{}
   { \label{fig:Kernel} 
A depiction of the scheme of the phase reconstruction process with an Asymmetric Pupil Fourier Wavefront sensor: (Top Left) phase map $\varphi$ within the asymmetric pupil mask, (Bottom Left) source image on the camera, (Bottom Right) phase $\Phi$ of the Fourier-plane signature in the (u,v) plane overlapped with half of the subaperture discretization, and (Top Right) the reconstructed pupil phase map $\hat{\varphi}$ from the pseudo-inverse matrix $\mathbf{A}^{+}$. The matrix A links the phase in the Fourier plane and in the pupil plane. }
\end{figure}

The kernel phase approach relies on building an accurate relationship between the phase of the Fourier transform of the focal plane image (Fourier phase) and the phase in the pupil~\citenum{martinache2010, martinache2013}. This can only be done accurately in the small aberration regime and requires a pupil asymmetry, be it phase or amplitude to break the degeneracy so the problem can be inverted and a unique solution found. This is typically achieved by using an asymmetric pupil mask that masks a small portion on one side of the pupil. The full process for sensing and extracting the phase aberration and associated correction can be seen in Fig.~\ref{fig:Kernel}.

Kernel phase techniques were originally used for imaging to study Brown dwarfs on the Palomar telescope~\citenum{pope2013}. In regards to wavefront control, the technique was experimentally validated on SCExAO in the H-band with the internal source first~\citenum{martinache2013}, and then on-sky where it was used to correct the first 10 Zernike modes~\citenum{martinache2016} and more recently to control the island effect (low-wind effect, atmospheric turbulence~\citenum{ndiaye2018}) in closed-loop operation. It operated successfully but was limited by a low temporal bandwidth and low sensitivity, both results of the properties of the detector used. 

%%%%%%%%%%%%%%%%%%%%%%%%%%%%%%%%%%%%%%%%%%%%%%%%%%%%%%%%%%%%%%%%%%%%%%%%%%%%%%%%%
\subsection{Phase shifting interferometry}
\label{subsec:psi1}
A phase shifting interferometer can be envisaged in numerous ways. Bottom et al. demonstrated one possible approach that worked in conjunction with the stellar double coronagraph in the Palm 3k instrument on the Palomar Telescope~\citenum{bottom2017}. The two vortex coronagraphs in series sent the rejected star light through the center of a reflective mirror, while the companion light and the residual leaked star light reflected from the mirror and were sent to the science camera. By placing a small mirror mounted on a linear translation stage inside the center of the Lyot stop (mirror that separated star from planet light), a small fraction of the rejected light was reflected back into the primary science beam and owing to its small size in the pupil diffracted into a large envelope in the focal plane which encompassed the entire PSF and halo interfering with it. By pistoning the small mirror the envelope was phase shifted with respect to the halo and hence the speckles were modulated in brightness. As the readout times of the detector were slow, the modulation in brightness revealed the quasi-static aberrations rather than speckles due to the turbulence (which evolve much faster). The brightness of the entire speckle halo was monitored simultaneously and CDI techniques were applied in post processing to extract the coherent and incoherent parts of the focal plane image and ultimately resulted in the detection of a known substellar companion~\citenum{bottom2017}. 

%%%%%%%%%%%%%%%%%%%%%%%%%%%%%%%%%%%%%%%%%%%%%%%%%%%%%%%%%%%%%%%%%%%%%%%%%%%%%%%%%
\subsection{Phase sorting interferometry}
\label{subsec:psi2}
\begin{figure}[ht]
    \centering
    \includegraphics[width=0.5\textwidth]{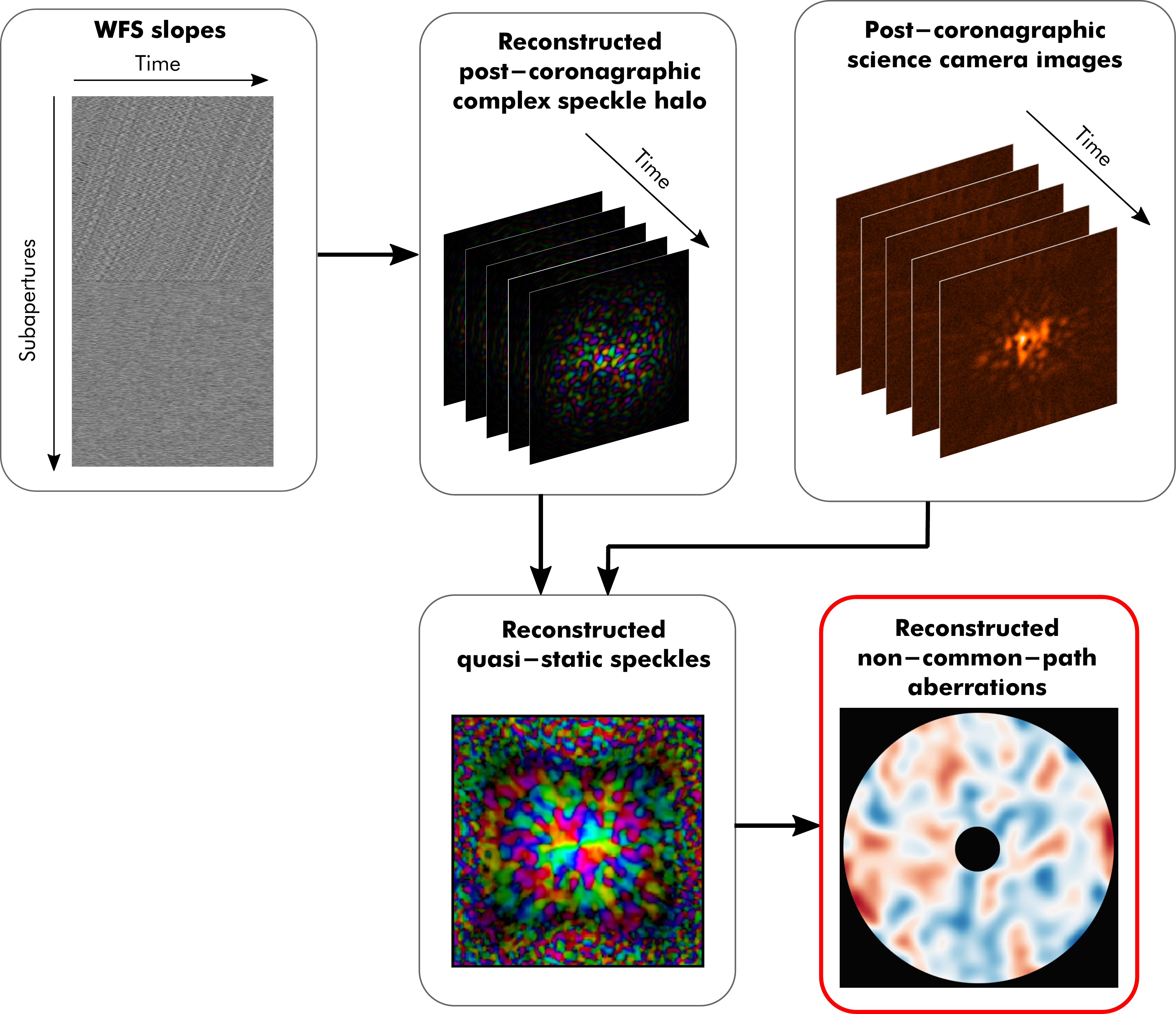}
    \caption{An overview of the measurement process of phase sorting interferometry.}
    \label{fig:PSI}
\end{figure}

Phase sorting interferometry is a non-invasive technique that can recover both phase and amplitude errors\citenum{Codona13}. It uses the residual wavefront errors of the adaptive optics system as phase diversity. The residual wavefront errors in an AO system are continuously measured by the wavefront sensor. The measured wavefront can be propagated analytically through the optical system, yielding the expected speckle halo in the science focal plane, both in phase and amplitude. These complex speckle estimates are then compared to the observed focal-plane images, and an estimate for the static speckle halo is extracted from the interferometric signal between residual AO speckles and unwanted static speckles. This estimate can be propagated back to the pupil plane and corrected by applying a wavefront sensor offset, if closed-loop operation is required. Additionally, an incoherent image can be computed using the same data set~\citenum{frazin2013,por2018}. An overview of the measurement process of phase sorting interferometry is shown in Fig.~\ref{fig:PSI}.

As phase sorting interferometry just requires a continuous stream of wavefront sensor telemetry and high-speed science camera images, and no modulation of the incoming wavefront, it can in principle be included on any existing high-contrast imaging system by only changing the control software. Phase sorting interferometry will be used for closed-loop NCP error correction on the high-contrast imaging arm on ERIS.

%%%%%%%%%%%%%%%%%%%%%%%%%%%%%%%%%%%%%%%%%%%%%%%%%%%%%%%%%%%%%%%%%%%%%%%%%%%%%%%%%%%%%%

\section{Lessons learned / Things to keep in mind}
From the forays into common-path WFSing to date, it is possible to suggest a few things to keep in mind when developing these systems. 
\begin{enumerate}
    \item If the goal is to apply these methods in real-time, the control system (RTC) must be flexible enough to support multiple sensors at real-time rates. Additionally, the availability of AO WFS telemetry is of high-value, to assist NCP WFS techniques and - later down the road - CDI data reduction efforts.
    \item The degree to which these sensors are effective depends upon the rate at which the speckles evolve and the rate at which they are sampled. Residual AO speckles have a life-time of approximately the ‘clearing time’ of the atmospheric phase screen over the input aperture (for a 10\,m telescope, and average wind speed of 10 m/s, the lifetime is roughly 1 second.) Of course, long-lived non-common path instrumental speckles can have a much longer time constant. Although these appear to be very reasonable time scales, most science cameras have an observing cadence that is much slower. For instance the HAWAII2 has typical full-frame readout times of 1.5\,s. The ideal science camera then is one that has low-noise readout with full frame readout rates of a few 10’s of Hertz. 
    \item Coherent sensing, by definition, requires that pathlength differences between the science and the reference (or probe) be long-term stable to within the coherence length $(\lambda * (\lambda/\delta \lambda))$. In particular for phase-modulated reference beams, the zero-point optical path length difference (OPD) must be maintained to within a fraction of a wavelength $(\lambda/8)$ over the full science observation. This requires careful opto/mechanical engineering. 
\end{enumerate}

\section{Milestones for the field}
The aim of this section is to outline some goals that would advance the field of common path wavefront sensing/CDI. The ultimate goal is to use the results of tests conducted to address the milestones laid out below in order to drive the design of future ELT and space mission high contrast imaging instruments. The goals outlined are not specific to a testbed or contrast level but are rather guidelines which should be used to direct characterization efforts. The milestones are posed as questions.

\begin{itemize}
    \item \textbf{How well do we know the complex field in the region where common-path WFSing/CDI is applied?} The ultimate goal here is to show that common-path WFSing techniques can be used to understand the complex field to a level of 90\% (or in other words the difference between our estimate of the incoherent field and the measured field is at the 10\% level). This goal can be broken up in tiers which include both laboratory and on-sky testing/validation. These should be tested in various contrast ranges (i.e.  lower contrasts for ground-based applications vs space-based applications). These tests can include standard CDI probing techniques like the introduction of known probe shapes/aberrations for estimation. The generation of incoherent speckles with the DM (modulated faster than the exposure time) may be useful for the ultimate testing of this milestone.
    
    The ultimate demonstration of the successful knowledge of the complex field can be best shown through a direct comparison of the contrast/detectability of a faint companion using standard high-contrast imaging techniques versus CDI techniques.
    
    \item \textbf{How well do we understand our instrument?} The goal is to develop a comprehensive model of an instrument and apply common path wavefront sensing/CDI techniques and determine if the experimental results are consistent with the model to within a factor of 2. This will tell us about any elements that were overlooked in the model, which will inform for future model design for similar systems.
    
    The second aspect of this goal is that once a comprehensive model has been established, it would be informative to study the gain of each common-path WFSing/CDI technique. This should be characterized as a function of: varying amounts of noise (or at least with realistic noise) for both the ground and space-based high contrast imaging cases.
    
    \item \textbf{What are the limits of my common path WFSing technique?} Ultimately, we want to understand the relative performance of each common-path WFSing/CDI technique in various regimes of wavefront correction/contrast. For each technique listed in table~\ref{tab:cpwfs} above, we want to understand the temporal, spectral and spatial bandwidths, science duty cycle and polarization state.
    
    The results of such characterizations will inform the appropriate parameter space within which each common-path WFSing technique is applicable and support appropriate selection for future ELT and space-based high contrast instruments. 
\end{itemize}

\section{Summary}
We have demonstrated that there are now many common path wavefront sensing/CDI techniques that are applicable to high contrast imaging. They are in various states of development and have shown differing levels of performance. As the environmental parameters for each testbed and demonstration varied wildly it is hard to make overarching conclusions on the most effective approach at this time, but it is clear that further development and direct comparisons will be needed. These techniques show great promise and will no doubt contribute to imaging extremely faint exoplanets on future GSMTs or space telescopes. We aim to expand this review and submit it as part of a series of publications to JATIS.

\acknowledgments % equivalent to \section*{ACKNOWLEDGMENTS}       
 
The authors would like to acknowledge the Lorentz Center for hosting and to a large extend funding the Optimal Optical Coronagraph workshop in Leiden 2017. Additional funding for the workshop was provided by the European Research Council under ERC Starting Grant agreement 678194 (FALCONER) granted to Frans Snik. This formed the platform where this work was carried out. 

% References
\bibliography{report} % bibliography data in report.bib
\bibliographystyle{spiebib} % makes bibtex use spiebib.bst

\end{document}